\documentclass[runningheads]{llncs}
\usepackage{graphicx}
\usepackage{hyperref}
\usepackage{xspace}
\usepackage{xcolor}
\usepackage{graphicx}
\usepackage[normalem]{ulem} % normalem do not replace \emph to underline
\usepackage{amsmath}
\usepackage{amssymb}
\usepackage{enumitem}
\usepackage{microtype}
\usepackage{hyphenat}
\usepackage{zed-csp}
\usepackage{mathtools}

\usepackage{hyperref}
\usepackage[anythingbreaks]{breakurl}
\usepackage[capitalize]{cleveref}
\crefname{section}{Sect.}{Sects.}
\Crefname{section}{Section}{Sections}

\usepackage{booktabs}
\usepackage{multirow}
\usepackage{makecell}
\usepackage{ragged2e}
\usepackage{longtable}
\usepackage{rotating}
\usepackage{lscape}
\usepackage{graphics}

\usepackage{subcaption}
\usepackage{wrapfig}

\usepackage{listings}

\usepackage[many]{tcolorbox}
\usepackage{lipsum}

\usepackage{bussproofs}
\usepackage{stackengine}

\usepackage{pifont}
\usepackage{tikz}
\usetikzlibrary{calc}
\usetikzlibrary{shapes.geometric}
\usetikzlibrary{decorations.pathreplacing}
\usetikzlibrary{positioning}
\usetikzlibrary{calc}
\usetikzlibrary{arrows}
\usetikzlibrary{shadows}
\usepackage{datetime} % for getting current date time

\newcommand*\circled[1]{\tikz[baseline=(char.base)]{
            \node[shape=circle,draw,inner sep=0.9pt] (char) {#1};}}

\newcommand{\XSpace}[1]{}

\newcommand{\yi}[1]{\textcolor{purple}{Yi: #1}}
\newcommand{\pt}[1]{\textcolor{violet}{Palina: #1}}

\newcommand{\Fix}[1]{\textcolor{red}{#1}}

\def\bitcoinA{%
  \leavevmode
  \vtop{\offinterlineskip %\bfseries
    \setbox0=\hbox{B}%
    \setbox2=\hbox to\wd0{\hfil\hskip-.03em
    \vrule height .3ex width .15ex\hskip .08em
    \vrule height .3ex width .15ex\hfil}
    \vbox{\copy2\box0}\box2}}

\newcommand{\DefMacro}[2]{\expandafter\newcommand\csname rmk-#1\endcsname{#2}}
\newcommand{\UseMacro}[1]{\csname rmk-#1\endcsname}

\newcommand{\InputWithSpace}[1]{\bgroup\def\arraystretch{1.1}\input{#1}\egroup}
\newcommand{\Code}[1]{{\ifmmode{\mathtt{#1}}\else$\mathtt{#1}$\fi}}
\newcommand{\CodeIn}[1]{{\ifmmode{\mathtt{#1}}\else$\mathtt{#1}$\fi}}

\newcommand{\cspsharp}{CSP\#\xspace}

\definecolor{silver}{RGB}{192,192,192}

\newcolumntype{R}[1]{>{\RaggedLeft\arraybackslash}p{#1}}
\newcolumntype{L}[1]{>{\RaggedRight\arraybackslash}p{#1}}
\newcolumntype{C}[1]{>{\centering\arraybackslash}p{#1}}

\newcolumntype{?}{!{\vrule width 1.3pt}}

\lstdefinestyle{tafile}{%
    basicstyle=\scriptsize\ttfamily,
    frame=single,
    rulecolor=\color{black},
    morekeywords={\$INHERIT, Update, Insert, Delete, reference, contain, call, HunkDep, Coverage},
    tabsize=2,
    numbers=left,                    % where to put the line-numbers; possible values are (none, left, right)
    numbersep=5pt,                   % how far the line-numbers are from the code
    stepnumber=1,
    numberstyle=\tiny\color{gray},   % the style that is used for the line-numbers
    keywordstyle=\sffamily,
}

\newtcolorbox{rqbox}[1]{%
    tikznode boxed title,
    enhanced,
    arc=0mm,
    boxrule=0.5pt,
    interior style={white},
    attach boxed title to top center= {yshift=-\tcboxedtitleheight/2},
    colbacktitle=white,coltitle=black,
    boxed title style={size=normal,colframe=white,boxrule=0pt},
    title=\textbf{Answer to }{\textbf{#1}}}

\hypersetup{final}

\begin{document}
\title{Formal Analysis of Composable DeFi Protocols}

\author{Palina Tolmach\inst{1,2} \and
Yi Li\inst{2} \and Shang-Wei Lin\inst{2} \and Yang Liu\inst{2}}

\authorrunning{Tolmach et al.}

\institute{Institute of High Performance Computing, Agency for Science, Technology and Research, Singapore \and
    Nanyang Technological University, Singapore \\
    \email{\{palina001,yi\_li,shang-wei.lin,yangliu\}@ntu.edu.sg}}

\maketitle              % typeset the header of the contribution
\begin{abstract}
Decentralized finance (DeFi) has become one of the most successful applications of blockchain and
smart contracts.
The DeFi ecosystem enables a wide range of crypto-financial activities, while the underlying smart
contracts often contain bugs, with many vulnerabilities arising from the unforeseen consequences of
composing DeFi protocols together.
In this paper, we propose a formal process-algebraic technique that models DeFi protocols in a
compositional manner to allow for efficient property verification.
We also conduct a case study to demonstrate the proposed approach in analyzing the composition of
two interacting DeFi protocols, namely, Curve and Compound.
Finally, we discuss how the proposed modeling and verification approach can be used to analyze
financial and security properties of interest.
\end{abstract}

\setcounter{footnote}{0}

\section{Introduction}\label{sec:intro}

With more than \$12 billions currently locked inside, decentralized finance (DeFi) becomes one of
the most prominent applications of the blockchain technology~\cite{defi-success}.
DeFi protocols implement various financial applications, including analogs of traditional-finance
use cases, such as lending~\cite{compound}, exchange~\cite{curve,UniswapWhitepaper},
investment~\cite{yearn}, etc.
These protocols give users access to digital assets, e.g., \emph{tokens}, and expose them to the
cryptocurrency market.
As an example, stablecoins are cryptocurrencies providing minimum volatility by pegging their
prices to fiat money, real-world commodity, or a more ``stable'' cryptocurrency, such as
ETH~\cite{moin2019stablecoins}.

At the same time, billions of dollars stored in DeFi stimulate the invention of new
security attacks.
% Within the last three months, more than
% three successful attacks were performed~\cite{bzx-attack,akropolis-attack,harvest-attack},
% including \$24M stolen from the Harvest protocol~\cite{harvest-attack}.
Unlike other smart contracts applications, the security of DeFi protocols
can be compromised by not only software vulnerabilities % such as reentrancy~\cite{akropolis-attack} and
% front-running~\cite{daian2019frontrunning},
but also unforeseen movements in the cryptocurrency market or arbitrage
and speculation opportunities. % such as pump-and-dump schemes~\cite{xu2019pump}.
For example, an attacker drained \$2M of funds from the (twice audited) Akropolis DeFi platform~\cite{akropolis-attack} through a well-studied reentrancy
vulnerability~\cite{Grossman2017,Liu2018Reguard,Samreen2020}.
% \pt{This paragraph was rewritten to better point out the problem:}
As another example, in March 2020, the network congestion caused by market instability led to major disruptions and
losses in some of DeFi protocols during the events of so-called ``Black
Thursday''~\cite{black-thursday}.

A distinctive feature of DeFi applications is their similarity to the pieces of so-called
\emph{Money Legos}~\cite{money-lego}.
In other words, the design of DeFi protocols
% do not operate in isolation from each other. Instead, they
often facilitates interoperability between them
% \yi{which applications?},
including the support of tokens issued by different DeFi platforms.
While the composability of DeFi applications enables the construction of a decentralized
financial ecosystem, integrations between protocols contribute to the creation
of new attack vectors.
For example, a recent attack on the Harvest yield aggregation protocol~\cite{harvest} was made
possible due to its dependence on the prices reported by the Curve decentralized exchange protocol~\cite{curve}.
%The attacker exploited the effect of impermanent loss, a characteristic issue of decentralized
%exchanges.
By performing a \$17M trade in Curve, the attacker could indirectly manipulate the price of tokens
in Harvest, obtaining \$24M of protocol funds~\cite{harvest-attack}.
% By modeling a set of composed DeFi protocols, we are able to reproduce similar attacks and ensure their detection by the framework under development.
% Furthermore, since DeFi protocols are linked together in an ecosystem of interacting applications,
% they are also prone to cascading failure, much like traditional finance.
An established way to rigorously verify correctness of safety-critical systems, including smart
contracts, is to employ formal analysis~\cite{woodcock2009formal}.
In the field of DeFi, security audits often involve formal analysis, but usually focusing only on
the verification of individual protocols.
Yet, the ``money-lego'' structure of the DeFi ecosystem demands compositional analysis, which
allows reasoning about the possible interplay between DeFi protocols and their impact on each
other.

To model and analyze the behaviors of composable DeFi protocols, we formulate general formal models
of components of DeFi protocols, particularly, \emph{tokens} and \emph{pools}. % The models are
%abstract and high-level.
Based on their actual implementations, we develop process-algebraic models of two widely used DeFi
protocols: a decentralized exchange---Curve Finance~\cite{curve}, and a lending
protocol---Compound~\cite{compound}.
In addition, we formally model the behavior of the USDC stablecoin.
    % To enable verification and property checking.
Using the developed model, we formally verify some of the (already stated) relevant properties of
the protocols under consideration.
Finally, we formulate safety and correctness properties that are expected to hold throughout the
interactions between the considered protocols. % Based on the
%example of the Curve Compound Pool.
% Finally, we provide a discussion on challenges and opportunities for compositional
% verification of interacting DeFi protocols.
\section{Background}\label{sec:back}
In this section, we provide necessary background for the rest of the paper.

\subsection{DeFi Protocols}\label{ssec:protocols}
We consider two common types of DeFi protocols: \emph{decentralized exchanges} (DEX)
and \emph{protocols for loanable funds} (PLF), a.k.a. \emph{lending protocols}.

\paragraph{Decentralized Exchanges}
%DEX, also called an \emph{automated market maker} (AMM),
DEX is one of the first and most popular DeFi applications.
While a centralized exchange has to match a seller with a specific buyer, a typical DEX uses smart
contracts to execute trades asynchronously~\cite{daian2019frontrunning,angeris2020uniswap}.
A \emph{pool}, implemented using smart contracts, stores the reserves of two or more types of
tokens and automatically determines the exchange rate between these tokens.
% A user willing to exchange a certain amount of tokens $\alpha$, supplies them to the pool and receives the amount
% of tokens $\beta$ in return.

A common way to determine the exchange rate between assets within a DEX pool is by maintaining a
\emph{constant-product} and/or \emph{constant-sum} invariant between the values of the tokens contained
in the pool.
Essentially, the invariant implies that if a user trades $t_1$ for $t_2$, the price of $t_1$
in the pool goes down, while the price of $t_2$ increases.
This model, therefore, provides an arbitrage opportunity for the users of DEXes, encouraging them
to deposit or sell tokens of type $t_2$ at a higher price, which thereby restores the balance between tokens.

\paragraph{Lending Protocols}
PLFs~\cite{gudgeon2020lending,perez2020liquidations} rely on smart contracts to mediate token
lending and borrowing.
Different from DEXes, lending pools collect assets of (usually) one token type from liquidity providers.
In return, the depositors are given pool tokens with the value constantly increasing from the
interest fees paid by borrowers.
The interest rate for borrowers depends on a chosen interest rate model and is usually decided by
the utilization rate---the ratio between the supply and demand of the pool.
To protect a protocol from the cryptocurrency volatility, the borrower is also supposed to supply a
collateral (e.g., in ETH or a stablecoin) that is bigger than the amount of borrowed funds by at
least a collateralization ratio.

\subsection{Formal Modeling and Verification}\label{ssec:formal-modeling}
% \Fix{TODO: insert some description of PAT and its components
% from~\cite{Sun2009PAT}.}

%\paragraph{CSP and CSP\#}
Communicating Sequential Process (CSP)~\cite{cspbook} is a formal language for describing
patterns of interaction for concurrent systems~\cite{roaw}.
A CSP model contains a set of synchronized or interleaving processes, each of which consists of a sequence
of ordered events.
For instance, a process $P$, with an event $a$ followed by another event $b$, can be written as
``$P = a \rightarrow b$''.
Multiple processes can be composed either sequentially or in parallel.
Sequential composition of two processes $P$ and $Q$ (denoted by $P;Q$) acts as $P$ first, and
acts as $Q$ upon the termination of $P$.
The two processes can also be composed in parallel and synchronized on an event $X$ ($P|[X]|Q$), or
asynchronously ($P|||Q$).
%Using external ($P\;\Box\; Q$) and internal ($P\sqcap Q$) choice operators, it is
%easy to describe the variations of the composed process behaviors.
Finally, a process $Q$ can interrupt another process $P$ when event $e$ happens ($P
\bigtriangledown e \rightarrow Q$).
The detailed syntax are summarized as follows.
{\small\[\begin{matrix*}[l]
  P := & \text{\quad} & STOP \text{\quad} % & (Process\;that\;do\;nothing) \\
    & | & SKIP % & (Process\;that\;terminates\;successfully) \\
    & | & e \rightarrow P % & (Prefixing)\\
    & | & P \;\Box\; P \\ % & (External \; choice)\\
    & | & P \sqcap P % & (Internal\; choice)\\
    & | & P \;\vert\vert\vert\; P % & (Interleaving) \\
    & | & P \;|[X]|\;P \text{\quad} % & (Synchronous \; parallel)\\
    & | & P \setminus X \\  % & (Hiding)\\
    & | & P;\; P  % & (Sequential \; composition)\\
    & | & \mathrm{if} \; b \; \mathrm{then} \; P\; \mathrm{else}\; P % & (Boolean \; condition)\\
    & | & P \;\triangledown\; P % & (Interrupt)
\end{matrix*}\]}%
\cspsharp~\cite{cspsharp,Liu2011pat} is an extension to CSP with embedding of data operations.
\cspsharp combines high-level compositional operators from process algebra with program-like codes,
which makes the language much more expressive.
The models and properties specified in \cspsharp can be checked using Process Analysis Toolkit
(PAT)~\cite{Sun2009PAT,Sun2008pat,Liu2011pat}, which is a framework for specification, simulation, and verification of
concurrent and real-time systems.
PAT supports event-based compositional models and efficient LTL model checking with various
fairness assumptions. Model checking~\cite{Clarke2000} is widely used to verify state-transition systems of one
or several interacting smart contracts against a temporal logic specification~\cite{tolmach2020survey}.
In this work, we use the model checker of PAT to verify the properties of individual and interacting
DeFi protocols, as described in~\cref{sec:evaluation}.

%Most importantly, PAT supports the verification of trace refinement
%checking~\cite{SunLD08b,WangS0LDWL12}, linear temporal logic verification with fairness
%assumption~\cite{SUNLDW08}, bounded model checking~\cite{SunLDS08,SunLDS08Journal}, fair model
%checking of parameterized systems~\cite{SUNLALD09},
%BDD-based discrete analysis of timed systems~\cite{tacas12} and assume-guarantee model
%checking~\cite{LADSL11,LADSL12}. % and Combining State Space Reductions with Global Fairness
%%Assumptions~\cite{ZhangSPLD11}

One unique feature of PAT is that it allows users to define static functions and data types
as C\# libraries.
These user-defined C\# libraries are built as DLL files and are loaded during execution, which compensates for the common deficiencies of model checkers on complex data operations and data types.
% \yi{edit: For instance, priority queue and set can be implemented
% to meet the need in the modeling of pricing functions.}
We utilize this capability and implement complex mathematical computations underlying the token
price calculation in C\#.
% We have also implemented C\# libraries which enable the storage and manipulation of larger integer values to improve the precision of our modeling approach.
Finally, the translation from high-level smart contract programming languages, such as Vyper and
Solidity, to C\# is straightforward.

\begin{comment}
The popularity of model checking is arguably caused by the suitability of both modeling and
specification formalisms to smart contracts description combined with the existence of established
automatic frameworks.
Although model checking is successful in verifying systems of several
smart contracts or users, its limitations are induced by the input language of a model
checker and the state explosion problem.
Support of diverse transition systems and temporal properties help
model checking capture different characteristics of smart contract execution, such as concurrency,
nondeterminism, or time constraints. Furthermore, with specifications
expressed in temporal logic, model checking is able to verify liveness properties and properties
of progress, e.g., liquidity.
Model checkers additionally verify conventional requirements of concurrent systems, such as deadlock- and livelock-freedom.
However, model-checking suffers from the state explosion problem, which requires the users to
apply abstraction techniques for smart contracts or assume a set of simplifications to its
execution. Model-checking approaches rarely consider the details of
smart contract execution on blockchain, such as the gas mechanism or a memory model.
\end{comment}

\section{Methodology}\label{sec:methodology}

%A recent survey~\cite{tolmach2020survey} on the existing verification techniques for smart
%contracts suggests that, in general, formal specification and verification approaches successfully
%provide a substantial level of security assurance.
%Since one of the key features of the proposed analysis framework is its ability
To reason about a system of interconnected protocols, we use a \emph{process-algebraic} approach
to model various components of the DeFi ecosystem.
First, we formally define the main components of DeFi applications along with the environment
models.
Then, we model two widely used Ethereum DeFi protocols and their interactions
using \cspsharp, by translating the major smart contract functions into CSP, in a similar fashion
to some of the previous work~\cite{Qu2018,Li2019bnb}.

\subsection{Protocol Formal Modeling}\label{ssec:encodings}

In this section, we propose formal definitions for the two key constituents of lending and exchange DeFi
protocols: \emph{token} and \emph{pool}.\footnote{Depending on the application, pools are also
referred to as \emph{markets}, \emph{vaults}, or \emph{pairs}.}
The behaviors of the aforementioned objects can be formalized as state transition systems,
and we focus on their states here.
% Functions that correspond to transitions can be found in the repository.
We leave the discussions on their transitions in~\cref{sec:appendix}.

We model the \emph{states} of users, smart contracts and the environment variables (e.g.,
\texttt{block.number}) as global variables in the \cspsharp model.
Functions, % and transitions \yi{or transactions?}
% \pt{transitions are supposed to be an equivalent of functions in our model, so I shouldn't have mentioned it here}
on the other hand, are translated into \emph{processes}.
% Events defined in the Vyper source code correspond to similarly-named and parametrized
% events in a CSP\# model.
Inspired by~\cite{bartoletti2020lending}, we assume a set of blockchain \emph{users}
% \footnote{In practice, a user is represented by an address on a blockchain.}
($\mathbb{U}$) and a set of \emph{tokens} ($\mathbb{T}$).
Tokens are programmable assets managed by smart contracts~\cite{Chen2020Token}.
The majority of tokens used in DeFi protocols, except the native platform cryptocurrency
ETH, are implemented in the form of a contract conforming to the ERC20 standard~\cite{ERC20}.
ERC20 regulates the development of fungible tokens by specifying the interface of the corresponding
smart contract, i.e., public functions and events that it should emit during executions.
In accordance with the standard, we define tokens in Definition~\ref{def:Token}.

\begin{definition}[Token] \label{def:Token}
A token $t \in \mathbb{T}$ is a tuple $(\mathbb{U}, TS, B, A, \mathbb{F})$, where $\mathbb{U}$ is a
set of users, $TS \in \mathbb Z_{\ge 0}$ is the total supply, $B: \mathbb{U} \mapsto \mathbb Z_{\ge
0}$ is the mapping from users to their token balances, $A: \mathbb{U}\times\mathbb{U} \mapsto
\mathbb Z_{\ge 0}$ specifies the allowances, i.e., amounts of token that a user is allowed to spend
from another user's balance, and $\mathbb{F}$ is the set of state-changing functions modifying the
state of the token.
\end{definition}%
%\Fix{(In reality, the mapping is from the user address $u$ instead of the user $U$.)}

Given a token $t \in \mathbb{T}$, we use $t.TS$ to denote its total supply and $t.A$ to denote its
allowances, and so on. The balance invariant of $t$ satisfies the formula:
$t.TS = \sum_{u \in \mathbb{U}} t.B(u)$.
$\mathbb{F}$ includes functions changing the values of $A$, $B$, and $TS$, e.g., \texttt{approve()},
\texttt{transfer()}, \texttt{transferFrom()}, \texttt{mint()}, \texttt{burn()}, etc.
\Cref{fig:token-impl} demonstrates a partial implementation of the state of the USDC token
in a model with $N$ participants.
Formally, the state of a user account $u$
% $S_U$, \yi{can we use $u$ and $U$ instead? it looks more intuitive}
is the set of balances of tokens in the user's possession, i.e., $\{t.B(u) \mid u \in \mathbb{U} \mbox{
and } t \in \mathbb{T} \}$.

%We abstract the set of functions of the token smart contracts as $\mathbb{F}$.
  % \pt{There are other parameters, such as Symbol and Decimals, but we do not consider them in the paper.}
  % \Fix{And define the corresponding transitions in Appendix.}

\begin{figure}[tb]
    \footnotesize\[\begin{matrix*}[l]
      \texttt{var USDC\_balances[N]:\{0..\} = [$b_{1}$,$b_{2}$,\dots,$b_{N}$];} \\
      \texttt{var USDC\_allowed[N][N]:\{0..\};} \\
      \texttt{var USDC\_totalSupply = $ts$;} \\
      \dots \\
      \end{matrix*}\]
      \caption{USDC token state implementation in CSP\#.}\label{fig:token-impl}
\end{figure}

\Cref{def:Pool} specifies pools, which are smart contracts used to aggregate a number of tokens.

\begin{definition}[Pool] \label{def:Pool}
A pool $P$ is a tuple $(\mathbb{T}_P, \mathbb{T}_P^R, \mathbb{F}_P)$, where
% $\mathbb{U}_P \subseteq \mathbb{U}$ is the \Fix{set of users participated in $P$},
$\mathbb{T}_P \subset \mathbb{T}$ is a set of pool tokens of $P$,\footnote{Most
of the pools in DeFi support a single pool token.} $\mathbb{T}_P^R \subset
\mathbb{T}$ is a set of liquidity tokens supported by $P$, and $\mathbb{F}_P$ is a set of functions $\{
(\mathbb{T}_P \times \mathbb{T}_P^R) \mapsto (\mathbb{T}_P
\times \mathbb{T}_P^R) \}_i$ changing the state of $P$.
\end{definition}

% \lsw{I think we don't need to define interest rate model or protocol parameters beause they can be abstracted in the pool changing functions}

% The set of tokens $\mathbb{T}_P^R \subset \mathbb{T}$ supported
% (i.e., stored) by $P$ is often referred to as \emph{liquidity} tokens.
% \emph{free tokens}~\cite{bartoletti2020lending}, sometimes also called reserves
Depending on the protocol application, liquidity tokens $\mathbb{T}_P^R \subset \mathbb{T}$
are used to facilitate decentralized token exchange, lending, investments, or other DeFi use cases.
Liquidity pools in DEX usually hold liquidity in two or more types of tokens~\cite{UniswapWhitepaper,curve,balancer},
while lending protocol~\cite{compound} or yield aggregator~\cite{yearn} pools typically accept a
single type of token as input.
In both cases, the users depositing tokens (a.k.a. \emph{liquidity providers})
% to earn passive income
receive a certain amount of pool tokens ($\mathbb{T}_{P}$), which represent user's share and can be
used to redeem the deposit with the earned interests from the pool.
$\mathbb{F}_P$ is a set of functions that can change the state of a pool.
\Cref{fig:pool-impl} illustrates the process-algebraic encoding of a state-changing
function that implements adding liquidity to a pool from the Curve protocol.
To mimic the atomic transaction execution model in Ethereum, we mark state-changing
processes as \emph{atomic}, so that their executions cannot be interrupted by an interleaving.

\begin{figure}[tb]
  \footnotesize\[\begin{matrix*}[l]
    \texttt{Curve\_addLiquidity(uamounts, min\_mint\_amount, sender) = atomic \{} \\
    \quad \dots \\
    % [\_uamounts[USDC] > 0]
    \quad \texttt{USDC\_transferFrom(user, curveDeposit, uamounts, \dots);} \\
    \quad \dots \\
    \quad \texttt{cUSDC\_mint(uamounts, curveDeposit);} \\
    \quad \dots \\
    \quad \texttt{cUSDC\_approve(curveSwap, cAmounts[USDC], curveDeposit);} \\
    \quad \texttt{Curve\_swap\_addLiquidity(cAmounts, min\_mint\_amount, sender);} \\
    \quad \texttt{cCrv\_transfer(user, cCrv\_mintAmount, curveDeposit);\}} \\
    % \dots \\
    \end{matrix*}\]
    \caption{CSP\# process for the \texttt{add\_liquidity()} function of a Curve pool.}
    \label{fig:pool-impl}
\end{figure}

\subsection{Protocol Composition}\label{ssec:composition}

Now, we illustrate how interactions between users and DeFi protocols (\emph{user-protocol}) as well
as interactions among different protocols (\emph{protocol-protocol}) can be modeled formally.
In both cases, the initiator of a transaction sends a certain amount of tokens to a receiving DeFi
protocol and/or receives some tokens from it in return.

In the case of \emph{user-protocol} interaction, we model the behavior of a user by a \emph{sequential composition} (denoted by `;') of one or more processes.
These processes correspond to the public state-changing functions of DeFi protocols and tokens
invoked by the user.
For example, the behavior of a depositor in Curve (i.e., \texttt{Curve\_Depositor}) is demonstrated
in~\cref{fig:user-composition}.
% We then define the behaviors of several users as a composition of processes corresponding to public
% state-changing functions of the protocols under consideration. \yi{what's the difference between
% several users and single user?}
% \pt{There is no difference, that was just my bad formulation. We define a set of users---each is
% a sequential composition of state-changing processes. Then, our system is an interleaving between
% these users}.

\begin{figure}[tb]
\footnotesize
\[\begin{matrix*}[l]
\texttt{Curve\_Depositor() = USDC\_approve(curveDeposit, suppliedTokens, user);} \\
\quad \texttt{Curve\_addLiquidity(suppliedTokens, minMintTokens, user);} \\
\quad \dots \\
\quad \texttt{cCrv\_approve(curveDeposit, add, user);} \\
\quad \texttt{Curve\_remove\_liquidity\_one\_coin(add, 0, user, true);}
\end{matrix*}\]
\caption{The implementation of Curve depositor behavior in CSP\#.}\label{fig:user-composition}
\end{figure}

The subject system is then modeled by an \emph{interleaving} (denoted by `{$\interleave$}') of such user
processes.
For instance, \cref{fig:interleaving-composition} shows the depositor, exchanger, and borrower
processes composed asynchronously, which simulates possible state changes in interacting protocols
caused by concurrently acting users.
% happening simultaneously; caused by any user behavior
The processes simulating state-changing functions are atomic, i.e., executing without interruption
so that the interleaving between user processes can only happen after a state-changing process is
finished.
%\pt{In other words, the simulation is a sequence of state-changing processes invoked by different
%users in (I guess) any possible order (allowed by sequential definition of the user process)}.
We simulate the block mining using a process that increases the value of the block number variable.

\begin{figure}[h!]
\footnotesize
\centering
% \[\begin{matrix*}[l]
\texttt{System() = Curve\_Depositor() $\interleave$ Curve\_Exchanger() $\interleave$ \\
  Compound\_Depositor() $\interleave$ Compound\_Borrower() $\interleave$ IncreaseBlockNum();}
% \end{matrix*}\]
\caption{The analyzed user composition.}\label{fig:interleaving-composition}
\end{figure}

% The user behaviors that involve the \emph{pool-pool} interactions (other, e.g. token transfers,
% could be parallel) are composed using an interleaving [] operator, reflecting the nondeterministic order or
% (atomic) transaction execution on Ethereum.

The \emph{protocol-protocol} interactions in DeFi smart contracts are external calls
to a function of another protocol.
Following a similar approach, we model smart contract functions with external calls to other DeFi
applications and token contracts as an \emph{atomic sequential composition} of corresponding
processes.
The sequential composition of two processes ensures that the former process has to finish before
the latter can start, so that the model operates similarly as the execution of internal
transactions in blockchain.
The CSP\# representation of a function that implements adding liquidity to a pool of the Curve DeFi
protocol is shown in~\cref{fig:pool-impl}.
The communication among users, tokens, and different protocols is simulated via shared global
variables, such as token balances shown in~\cref{fig:token-impl}.
% \yi{Can't see where shared global variables are.}
\section{Evaluation}\label{sec:evaluation}

\begin{figure}[tb!]
    \centering
    \includegraphics[width=0.9\textwidth]{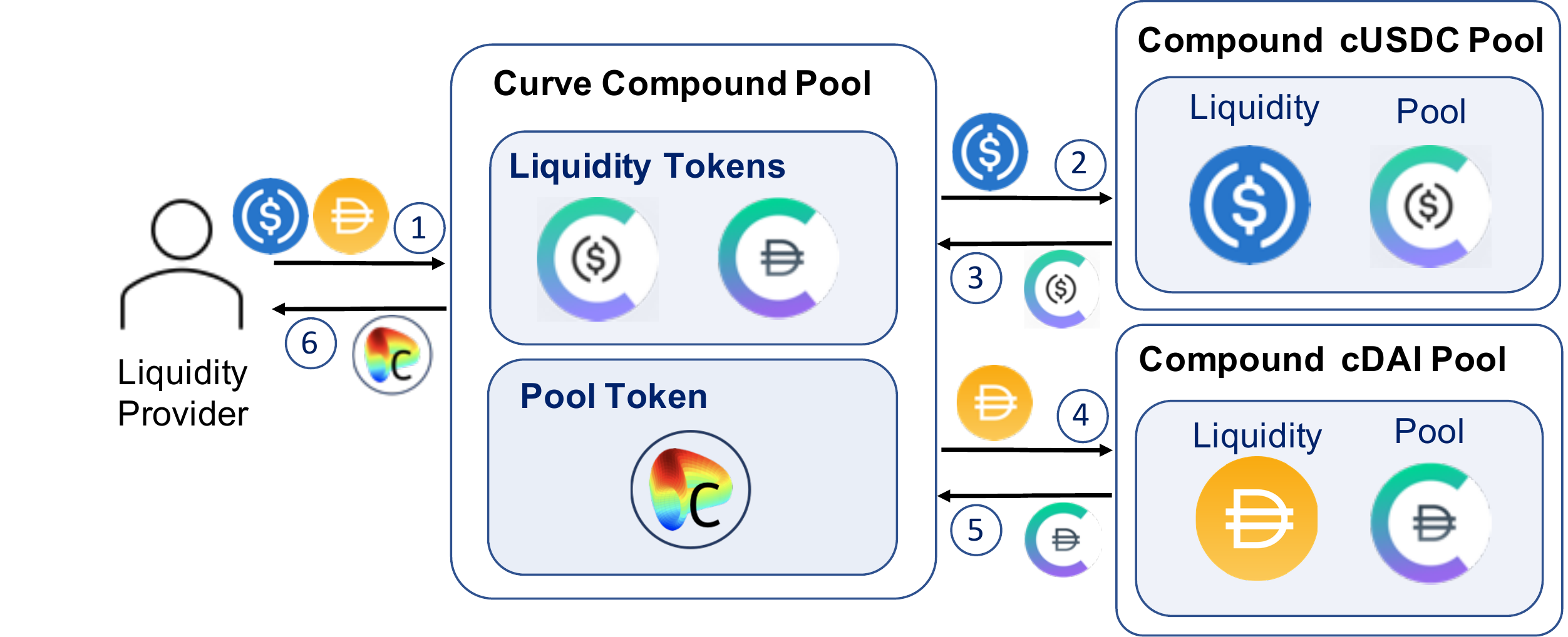}
    \caption{A scheme of token transfers between Curve Compound pool participants.}\label{fig:flow}
\end{figure}

%Based on the methodology described in~\cref{sec:methodology}, we modeled selected functionality
%of two interacting DeFi protocols in CSP\#.
In this section, we evaluate our modeling approach by checking a set of relevant properties on
Compound pool of the Curve DEX\footnote{https://www.curve.fi/compound} using PAT and report on the
results of property verification. We performed the evaluation on a virtual machine with
Windows 10, 8GB RAM and 1 CPU core, using PAT version 3.5.0. The virtual machine is running on
MacOS Catalina v.10.15.7, 32GB RAM and 2 GHz quad-core Intel Core i5 processor.

The Curve Compound pool allows trading between a pair of stablecoins: USDC and DAI.
Under the hood, the Curve pool transfers its USDC and DAI to a lending platform Compound, in
exchange for the corresponding Compound's pool tokens---cUSDC and cDAI.
cUSDC and cDAI are, therefore, used for all the operations within the Curve Compound pool.
\Cref{fig:flow} outlines the process of adding liquidity to the Curve Compound pool: \circled{1} a liquidity
provider sends USDC and/or DAI to the pool; \circled{2} Curve supplies the received USDC to the USDC Compound pool
and \circled{3} receives an appropriate amount of cUSDC in return; \circled{4}-\circled{5} the same process is repeated for DAI/cDAI;
\circled{6} the user receives a certain amount of cCrv---a pool token of the Curve Compound pool.
% \Cref{fig:flow} displays the token movements that are initiated by a user adding liquidity
% to the Curve Compound pool. \yi{Fig.5 needs more explanations than this: what are (1)-(5) in the
% figure? Can you walk through the process instead of the previous two sentences.}

State-changing actions of interest include providing and removing liquidity in both Curve and
Compound, exchanging tokens in Curve, and taking/repaying a loan in Compound.
In this paper, we mostly concentrate on the operations that involve USDC: in our model,
a liquidity provider on Curve adds and withdraws liquidity in USDC, while Compound
depositors and borrowers also perform the corresponding actions with the USDC Compound pool.
To model slippage and front-running that can occur in the pool of a DEX, the token exchanges
between cUSDC and cDAI in Curve can happen in both directions.
We assume that the modeled trading activity reflects the possible changes in the USDC/DAI exchange rate,
which we do not explicitly consider otherwise.
In addition, since we focus on the operations involving the USDC stablecoin, we simplify
the implementation of DAI to basic ERC20 functionality and do not consider the underlying stabilizing
mechanism implemented by MakerDAO.
We modeled pools and tokens by manually translating their source code written in Solidity or Vyper to
CSP\# and C\# languages supported by PAT. While the translation between high-level languages (e.g., Solidity/C\#) is straightforward,
data operations and programming constructs supported by CSP\# also facilitate translation to a modeling
language. The source code of the model can be found in a repository: \url{https://github.com/polinatolmach/DeFi-csp-models/}.

\begin{table}[tb]
  \begin{center}
    \caption{A summary of verified properties.}\label{tab:properties}
    \resizebox{\columnwidth}{!}{
    \begin{tabular}{L{0.5cm}L{2.8cm}L{8.3cm}ccc}
      \toprule
      \textbf{\#} & \textbf{Properties} & \textbf{LTL Formulae} & \textbf{Protocols}
      & \textbf{\makecell[c]{Results}} & \textbf{\makecell[c]{Stats}} \\
      \midrule
      (1) & Balance Invariants & \makecell[l]{\texttt{$\Box$((sum(cCrv\_balances) == cCrv\_totalSupply) \&\&}
        \\ \texttt{sum(cDAI\_accountTokens) == cDAI\_totalSupply)\dots)}}
        & \makecell[c]{All Tokens} & Valid & \makecell[r]{Time (s): 275.5 s\\ \#State:
        127337 \\ \#Transition: 133763} \\
      \midrule
      (2) & \makecell[l]{Proportional Token \\ Exchange} & \makecell[l]{\texttt{$\Box$((suppliedTokens > 0) $\rightarrow$} \\
      $\Diamond$\texttt{((mintedCTokens > 0) \&\& (mintedCCrvTokens > 0)))}}
        & \makecell[c]{Curve \\ Compound} & Valid & \makecell[r]{Time (s): 277.7 s \\
        \#State: 127367 \\ \#Transition: 133821}\\
      \midrule
      (3) & \makecell[l]{Non-decreasing \\ Exchange Rate} & \texttt{$\Box$(prevExchangeRate $\leq$ newExchangeRate)}
      & Compound & Valid & \makecell[r]{Time (s): 277.9s \\ \#State: 127337 \\
      \#Transition: 133763} \\
      \midrule
      (4) & Non-negative Profit & \texttt{$\Box$ (Mint.cUSDC $\rightarrow$ $\Box$(depositorProfit $\geq$ 0))}
        & Compound & Invalid & \makecell[r]{Time (s): 1.0 s \\ \#State: 430 \\
        \#Transition: 453} \\
      \midrule
      (5) & Bounded Loss & \makecell[l]{\texttt{$\Box$ (AddLiquidity $\rightarrow$}
        \\ \texttt{$\Box$(depositorLoss $\leq$ ADMISSIBLE\_LOSS)}}
        & Curve & Invalid & \makecell[r]{Time (s): 0.5s \\ \#State: 177
        \\ \#Transition: 196} \\
      \bottomrule
    \end{tabular}}
  \end{center}
\end{table}

Based on the defined model, we formulated and verified properties for tokens, individual DEX and
lending DeFi applications as well as their composition.
LTL formulae and verification results for the properties are demonstrated in~\cref{tab:properties}.
The first property in~\cref{tab:properties} is the \emph{Balance Invariant}~\cite{tolmach2020survey}---an important property related to
tokens, which we verify for all the tokens involved in the modeled composition: stablecoins (USDC and DAI)
and pool tokens (cCrv, cUSDC, etc.).
% \pt{Should I add a citation to what a balance invariant is?}
Property (2) in~\cref{tab:properties} is a token-related requirement for a composition of protocols stating that
%  should hold throughout the interaction between different protocols states that
\emph{the positively-valued tokens should never produce zero tokens}
(Proportional Token Exchange)~\cite{Bernardi2020}.
We verified that this requirement holds for all pairs of tokens involved in the process of adding
liquidity to the Curve Compound pool~(\cref{fig:flow}).

Among the properties of individual protocols, our model allows verification of \emph{the exchange rate of
the pool token in Compound being non-decreasing}, meaning that a liquidity provider always receives
a guaranteed interest on her deposit (Property (3) in~\cref{tab:properties}). % This property has been formulated by Bartoletti et
%al.~\cite{bartoletti2020lending}.
For a liquidity provider on Compound, we additionally checked whether her \emph{profit from
providing and then redeeming liquidity can only be non-negative} (Property (4) in~\cref{tab:properties}).
While this requirement holds under normal conditions, it does fail in the event of
\emph{overutilization}, i.e., if the pool does not have enough liquidity to repay the depositor.
% Since the user can't redeem the original token due to illiquidity.
To model \emph{overutilization}, we defined a user who borrows all the available liquidity from the
Compound pool.
For simplicity, we omited the collateralization requirements in our model---each loan is assumed to
be collateralized using the token that is not considered in the current model (e.g., ETH).
% \Fix{For Curve, there is a protocol-specific requirement stating that the value of $D$ should never decrease
% unless pool parameters are changed.}
Although the simplifications assumed in our model allow reaching overutilization easier than it is in reality,
it % ``it'' is overutilization
remains one of the main risks associated with lending protocols~\cite{bartoletti2020lending}.

Overutilization in a Compound pool causes a violation of an analogous property defined for a Curve liquidity
provider, showing the potentially harmful effects of composability. In other words, the users of both Compound
and Curve are not always able to redeem their original deposit back. Considering that a liquidity provider in a DEX
can legitimately suffer losses from the \emph{impermanent loss}, % should I insert a reference to what IL is? Should I use another word?
the property (5) in~\cref{tab:properties} requires \emph{the loss to be bounded by a certain value}, which we set to 20\%
of the original deposit. % that can be an arbitrary number
This requirement can also be violated in an anticipated way due to \emph{front-running} and
\emph{slippage} caused by massive trades made by other users.
% In this way, the user can't redeem his token too.
The violations of both properties are identified by PAT in sub-second time.
Being an on-the-fly model checker, PAT stops constructing and exploring the state space after
detecting the violation, which explains the time discrepancy between the verification of
properties (1)--(3) and (4), (5). For both violated properties, the reachability analysis in PAT also helps identify the maximum
possible losses and profits for both Curve and Compound depositors.
Finally, we confirmed the violation of properties on a locally deployed Ethereum network,
assuming the same set of simplifications to smart contracts as in the model.

% ERC20 standards involves implementation of the following functions:
% transfer(address recipient, uint256 amount), transferFrom(), approve(address recipient, uint256 amount) -- all these functions return bool
% getters: totalSupply(), balanceOf(address account), allowance(address owner, address spender);
% They emit Transfer(from,to,value), Approve(owner,spender,value);

% For the Curve pool, the successful Deposit and Exchange can only execute after an Approve event;
% Remove_Liqudity/Redeem (in Curve) can only happen if there is Deposit/Mint before; Burn can only happen after a Mint or a Transfer, etc.;
% Infinite mint is not possible; Mint is preceded by a Transfer of an underlying token;
% In Compound, accrueInterest happens infinitely often;

% The model formalizes the interaction between two widely used DeFi protocols, namely, Curve~\cite{curve} and Compound~\cite{compound}.
% The conceptual model allows verification of a multi-contract invariant that should hold throughout the lifetime of the involved smart contracts.
% However, future steps in the framework development include an extension to PAT with custom arithmetical types for a more realistic and practical model.
% Other planned additions include implementation of Ethereum-specific built-in functions, such as \emph{delegate calls}.
% Verification of large systems may cause state explosion, in which case we might want to look at the compositional verification and assume-guarantee reasoning.

The performed evaluation demonstrates the suitability of applying process algebra CSP for modeling
concurrently acting users and DeFi protocols on blockchain.
The results also confirm that model checking can automatically reveal undesirable conditions in the
operation of a single DeFi protocol or a composition of those.
However, with expanding the composition of modeled users and protocols,
the number of states grows exponentially.
To combat the state explosion problem, we consider utilizing techniques from the area of compositional verification,
such as assume-guarantee reasoning~\cite{Lin2014compositional,Lin2012compositional,Lin2015compositional}, which we leave for future work.
\section{Related Work}\label{ssec:related-work}

The analysis of DeFi protocols is a relatively new field.
The existing works often focus on specific types of DeFi protocols or investigate abnormal
behaviors observed in the wild.
%done on empirical analysis of certain aspects of DeFi operation, such as flash
%loans~\cite{wang2020flashloans}, oracles~\cite{liu2020oracles}, pump-and-dump
%activities~\cite{xu2019pump}, stablecoins~\cite{KlagesMundt2020stablecoins},
%or the operation and interest rate models of lending protocols~\cite{gudgeon2020lending}.
For example, Liu and Szalachowski explored the usage of \emph{oracles} in four major DeFi
platforms~\cite{liu2020oracles}, revealing the operational issues inherent in oracles and
common deviations  between the real and reported prices.
% Their large-scale empirical analysis of oracle transactions reveals common deviations of reported
% prices from the real ones, as well as operational issues inherent in oracles, even though
% they are often treated as trusted parties by the protocols.
% \cite{xu2019pump} also belongs here

A number of articles analyze the attack vectors that involve a
\emph{flash loan}~\cite{qin2020attackingFL,gudgeon2020crisis}, while
Wang et al.~\cite{wang2020flashloans} proposed a framework that allows the identification and
classification of flash loan transactions.
Their technique is able to detect speculative usage of flash loans and other potentially harmful
behaviors.
% Qin et al.~\cite{qin2020attackingFL} analyzed the attack vectors involving a \emph{flash loan}, while
% Furthermore, Gudgeon et al.~\cite{gudgeon2020crisis} showed how flash loans can be used to attack
% the governance system of a DeFi protocol and proposed a stress-testing framework for DeFi lending
% protocol to estimate its risks of becoming undercollateralized.
% There is also a related paper on pump-and-dump activities~\cite{xu2019pump}.

Several studies explore the operation and properties of DeFi
\emph{lending protocols}~\cite{gudgeon2020lending,kao2020compound,perez2020liquidations,bartoletti2020lending}.
% The authors of~\cite{gudgeon2020lending} investigated liquidity, interest rates, and market
% efficiency of such lending protocols.
% Among other findings, the study suggests that the borrowing rates observed in the protocols are
% interdependent, with the periods of high utilization often shared across protocols.
% TODO: Maybe discuss the work on quantifying returns of DEX liquidity
% providers~\cite{evans2020liquidity}
Kao et al.~\cite{kao2020compound} utilized agent-based simulations to analyze the market risks
faced by the Compound lending protocol users.
Stress-tests were performed to demonstrate the scalability of the protocol on a larger borrow size
under reasonably volatile conditions.
Formal models of lending protocols and their pools were formulated in two recent
publications~\cite{perez2020liquidations,bartoletti2020lending}.
Bartoletti et al.~\cite{bartoletti2020lending} also formulated the fundamental properties
of lending pools and typical ways of their interaction with other DeFi protocols.
Meanwhile, Perez et al.~\cite{perez2020liquidations} utilized the abstract formal model of Compound
to explore the possibility of liquidations of undercollateralized positions.
Different from the discussed publications, this paper formulates a more general formal model of a
pool, which can be used to formalize both lending and DEX protocols.

\begin{comment}
Front-running~\cite{daian2019frontrunning} is a well-known security issue of DeFi applications, and
DEX in particular.
To study the risks caused by transaction-ordering dependencies in Ethereum, Daian et
al.~\cite{daian2019frontrunning} formally model behaviors of bots engaged in arbitrage and
front-running activities.
\end{comment}
In addition, Klages-Mundt et al.~\cite{KlagesMundt2020stablecoins} proposed a framework for
modeling and classifying stablecoins. The authors also formulated and examined the risks associated
with stablecoins and their use in the DeFi ecosystem.
The formal model of a token considered in this paper is of a higher level and does not cover its
underlying economical mechanism.

Finally, a recent publication by Bernardi et al.~\cite{Bernardi2020} proposed a set of invariants
that are relevant for individual DeFi protocols, including DEXes and lending platforms.
While our study involves verification of some of the invariants proposed in this article,
we further extend them to the system of interacting DeFi protocols.
%  as well as complement with other relevant properties discussed in~\cref{ssec:properties}.
% (e.g., ``proportional token distribution'' and
% ``aggregated ledger integrity'')
\section{Conclusion and Future Work}\label{sec:conclusion}

% \Fix{\subsection{Assumptions and Limitations}}

In this paper, we proposed formal definitions for the main components of DeFi protocols and an
approach to model their implementations and interactions in a process-algebraic modeling language.
We demonstrated how model checking can automatically verify correctness properties for a
composition of DeFi protocols and tokens.
The proposed technique successfully identifies the DeFi-specific conditions
% (e.g., slippage and overutilization)
that cause the violations of these properties.

As future work, we plan to enrich the models to account for functionality related to
liquidity-mining and governance mechanisms in the considered DeFi protocols.
We would also like to extend the set of properties to cover both security vulnerabilities and the
cryptoeconomical aspects of DeFi executions. Finally, to address the state explosion problem,
we plan to integrate techniques from the area of compositional verification.

\subsubsection*{Acknowledgements.}
This research is partially supported by the Ministry of Education, Singapore, under
its Academic Research Fund Tier 1 (Award No. 2018-T1-002-069) and Tier 2 (Award No. MOE2018-T2-1-068),
and by the National Research Foundation, Singapore, and the Energy Market Authority, under its Energy Programme
(EP Award No. NRF2017EWT-EP003-023). Any opinions, findings and conclusions or recommendations expressed in this material
are those of the authors and do not reflect the views of National Research Foundation, Singapore and the Energy Market Authority.

\bibliographystyle{splncs04}
\bibliography{references}

\appendix
\section{Model Transition Implementation}\label{sec:appendix}

In this section, we discuss some of the processes that correspond to state-changing functions
of the protocols and tokens under consideration.
As described in \Cref{ssec:encodings}, we model key components of DeFi protocols, such as token and
pool smart contracts, as state transition systems.
States of the protocols are mostly defined by the values of smart contract variables, % e.g.,
%\texttt{balances}
while transitions correspond to the state-changing functions of the smart contracts.

\subsection*{Token Functions}

\begin{figure}[h]
\centering
\begin{subfigure}[b]{1.01\textwidth}
\centering
\includegraphics[width=\linewidth]{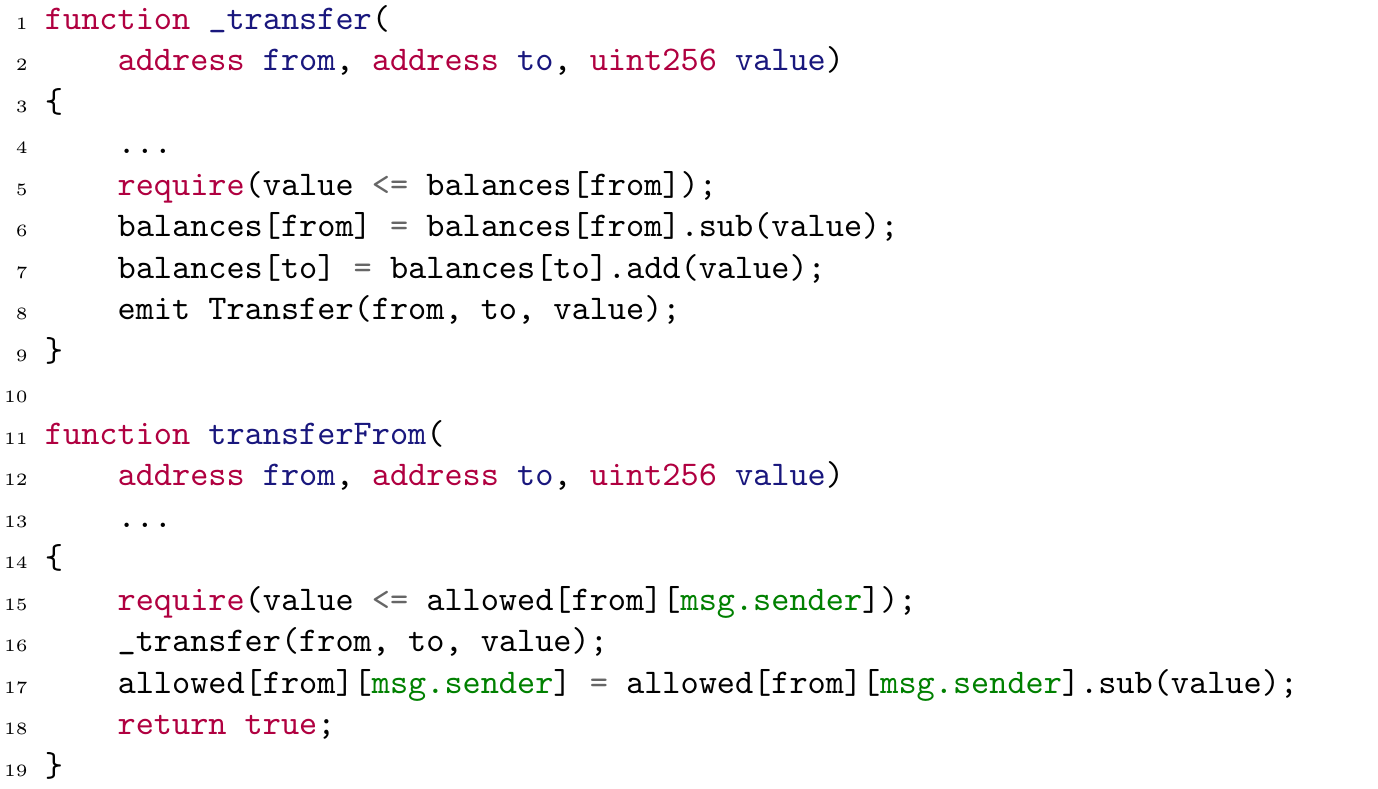}
\caption{Solidity implementation of \texttt{transfer()} and \texttt{transferFrom()} functions.}
\label{fig:usdc-solidity}
\end{subfigure}
\begin{subfigure}[b]{\textwidth}
\centering
\footnotesize
\begin{flalign*}
& \texttt{USDC\_transfer(to, value, from) = atomic \{} &\\
& \quad \texttt{if (value <= USDC\_balances[from]) \{} &\\
&\qquad \texttt{transfer.from.to.value\{} &\\
& \qquad\quad \texttt{USDC\_balances[from] -= value;} &\\
&\qquad\quad \texttt{USDC\_balances[to] += value;\} -> Skip\}} &\\
&\quad \texttt{else \{REVERT -> Reverting()\}\};} &\\
\\
& \texttt{USDC\_transferFrom(from, to, value, sender) = atomic \{} &\\
& \quad \texttt{if (value <= USDC\_allowed[from][sender]) \{} &\\
&\qquad \texttt{USDC\_transfer(to, value, from);}\\
&\qquad \texttt{tau\{USDC\_allowed[from][sender] -= value;\} -> Skip\}}  &\\
&\quad \texttt{else \{REVERT -> Reverting()\}\};} &
\end{flalign*}
\caption{CSP\# definition of \texttt{transfer()} and \texttt{transferFrom()} functions.}
\label{fig:usdc-csp}
\end{subfigure}
\caption{Solidity and CSP\# implementations of functions in USDC token.}
\label{fig:usdc-func}
\end{figure}

In a token implementation, state-changing functions are usually concerned with updating the values
of variables that track the token amounts and allowances.
\Cref{fig:usdc-solidity} contains the original implementation of \texttt{transfer()} and
\texttt{transferFrom()} functions in the USDC stablecoin smart
contract.\footnote{https://etherscan.io/address/0xA0b86991c6218b36c1d19D4a2e9Eb0cE3606eB48}
\Cref{fig:usdc-csp} illustrates the definition of a corresponding process in CSP\# used in our
model of USDC.
The process changes the state of some of the involved shared variables shown
in~\cref{fig:token-impl}, such as \texttt{USDC\_balances}.

\subsection*{Compound Pool Functions}

\begin{figure}[h]
\centering
\begin{subfigure}[b]{1.01\textwidth}
\centering
\includegraphics[width=\linewidth]{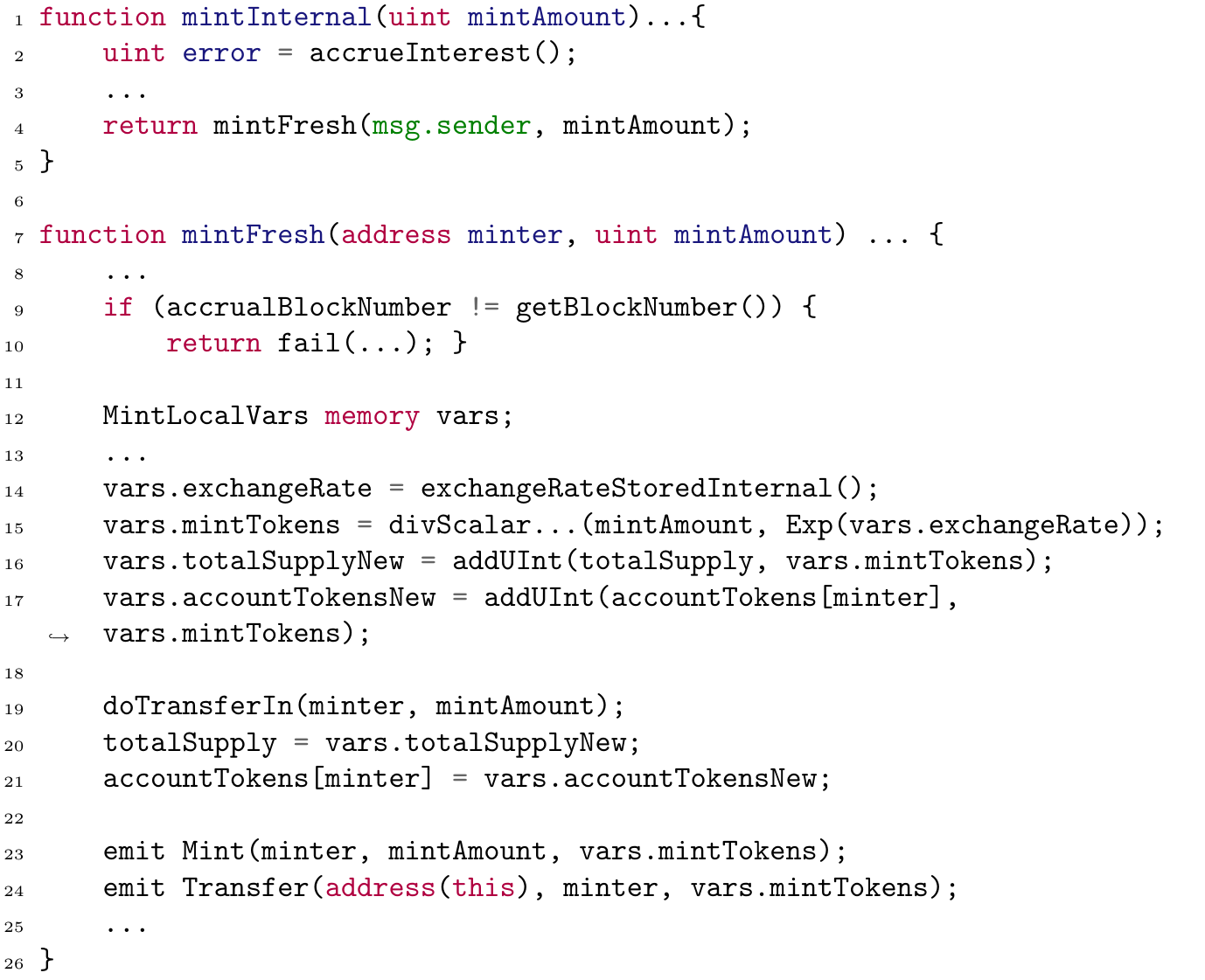}
\caption{Simplified Solidity code of cUSDC \texttt{mint()} function.}
\label{fig:mint-solidity}
\end{subfigure}
\begin{subfigure}[b]{1\textwidth}
\centering
\footnotesize
\begin{flalign*}
& \texttt{cUSDC\_mint(mintAmount, sender) = atomic \{ } &\\
& \quad \texttt{cUSDC\_accrueInterest();} &\\
& \quad \texttt{cUSDC\_mintFresh(sender, mintAmount)\};} &\\
\\
& \texttt{cUSDC\_mintFresh(minter, mintAmount) = \{ } &\\
&\quad \texttt{(if (accrualBlockNumber != currentBlockNumber) \{} &\\
&\qquad\quad \texttt{REVERT -> Reverting()\}} &\\
&\quad \texttt{else \{cUSDC\_exchangeRateStored(); } &\\
&\qquad \texttt{tau\{mintTokens = call(calcMintCUSDC, mintAmount, exchangeRates);} &\\
&\qquad\qquad \texttt{cUSDC\_totalSupply += mintTokens;} &\\
&\qquad\qquad \texttt{cUSDC\_accountTokens[minter] += mintTokens;...\} -> } &\\
&\qquad \texttt{USDC\_transferFrom(minter, compCUSDC, mintAmount, compCUSDC);} &\\
&\qquad \texttt{Mint.cUSDC -> mint.minter.mintAmount.mintTokens ->} &\\
&\qquad \texttt{transfer.compCUSDC.minter.mintTokens -> Skip\})\};} &
\end{flalign*}
\caption{CSP\# definition of cUSDC \texttt{mint()} function.}
\label{fig:mint-csp}
\end{subfigure}
\caption{Implementations of \texttt{mint()} function in Compound cUSDC pool.}
\label{fig:mint-func}
\end{figure}

\begin{figure}[h]
\centering
\begin{subfigure}[b]{1.01\textwidth}
\centering
\includegraphics[width=\linewidth]{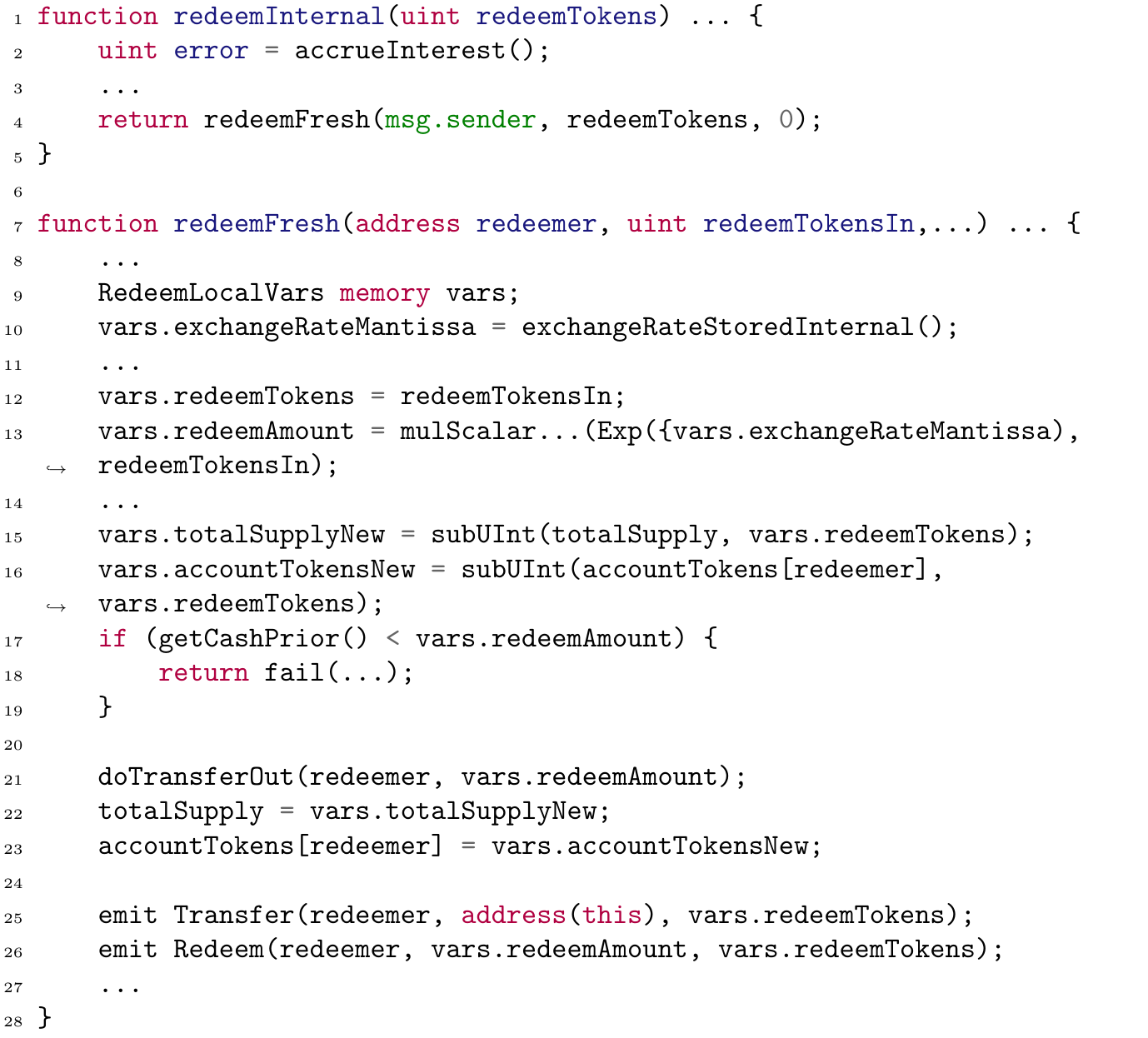}
\caption{Simplified Solidity code of cUSDC \texttt{redeem()} function.}
\label{fig:redeem-solidity}
\end{subfigure}
\begin{subfigure}[b]{1\textwidth}
\centering
\footnotesize
\begin{flalign*}
& \texttt{cUSDC\_redeem(redeemTokensIn, sender) = atomic \{ } &\\
& \quad \texttt{cUSDC\_accrueInterest();} &\\
& \quad \texttt{cUSDC\_redeemFresh(sender, redeemTokensIn, 0)\};} &\\
\\
& \texttt{cUSDC\_redeemFresh(redeemer, redeemTokensIn, redeemAmountIn) = \{ } &\\
&\quad \texttt{cUSDC\_exchangeRateStored(); } &\\
&\quad \texttt{tau\{redeemTokens = redeemTokensIn;} &\\
&\quad\qquad \texttt{redeemAmount = call(calcRedeemCUSDC, redeemTokensIn, exchangeRates);\} ->} &\\
&\quad \texttt{(if (USDC\_balances[compCUSDC] <= redeemAmount) \{} &\\
&\qquad\quad \texttt{REVERT -> Reverting()\}} &\\
&\quad \texttt{else \{USDC\_transfer(redeemer, redeemAmount, compCUSDC); } &\\
&\qquad\quad \texttt{tau\{cUSDC\_totalSupply += redeemTokens;} &\\
&\qquad\qquad \texttt{cUSDC\_accountTokens[redeemer] -= redeemTokens;...\} ->} &\\
&\qquad \texttt{transfer.redeemer.redeemTokens ->} &\\
&\qquad \texttt{redeem.redeemer.compCUSDC.redeemAmount -> Skip\})\};} &
\end{flalign*}
\caption{CSP\# definition of cUSDC \texttt{redeem()} function.}
\label{fig:redeem-csp}
\end{subfigure}
\caption{Implementations of \texttt{redeem()} function in Compound cUSDC pool.}
\label{fig:redeem-func}
\end{figure}

The modeled functionality of a pool in the Compound protocol includes depositing and redeeming liquidity
(e.g, USDC) and taking or repaying a loan.
\Cref{fig:mint-func,fig:redeem-func} demonstrate the simplified Solidity code and CSP\# definitions for
functions that realize minting and redeeming of cUSDC tokens\footnote{https://etherscan.io/address/0x39aa39c021dfbae8fac545936693ac917d5e7563}
(\texttt{mint()} and \texttt{redeem()}, respectively).
In Compound, the same smart contract implements both token and pool functionality, therefore, these functions also correspond to
providing and redeeming liquidity from the Compound USDC pool, where cUSDC serves as a pool token.
Following the definition of the pool state given in~\cref{ssec:encodings},
the processes and functions shown in \Cref{fig:mint-func,fig:redeem-func} change the state of a liquidity token (USDC) and a pool
token (cUSDC). The state of a liquidity token is changed through its transfers (\texttt{USDC\_transfer()})
and of a pool token---via updating \texttt{cUSDC\_totalSupply} and \texttt{cUSDC\_accountTokens} variables.

\subsection*{Curve Pool Functions}

\begin{figure}[h]
\centering
\begin{subfigure}[b]{1\textwidth}
\centering
\includegraphics[width=\textwidth]{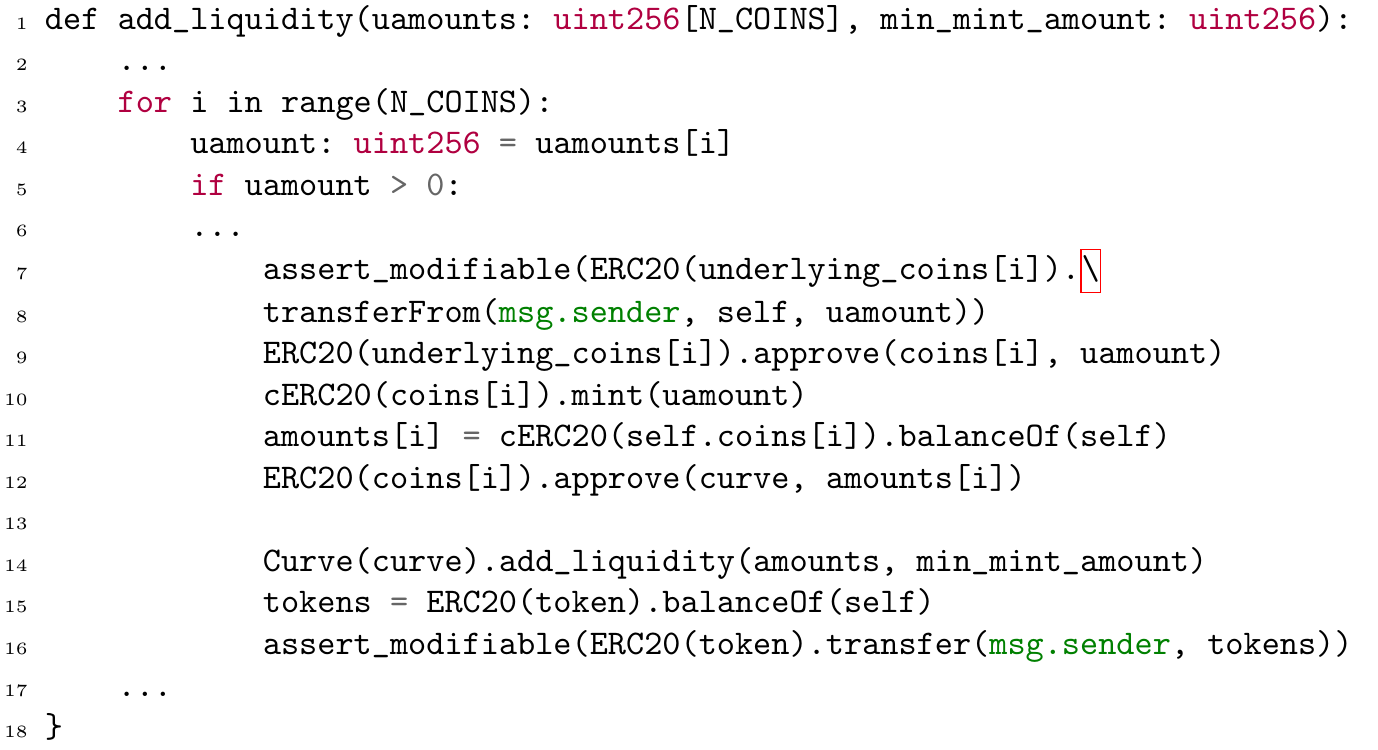}
\caption{Simplified Vyper implementation of \texttt{add\_liquidity()} function.}
\label{fig:add-vyper}
\end{subfigure}
\begin{subfigure}[b]{1\textwidth}
\centering
\footnotesize
\begin{flalign*}
&\texttt{Curve\_addLiquidity(uamounts, min\_mint\_amount, sender) = } &\\
&\quad \texttt{atomic \{[uamounts > 0] } &\\
&\quad \dots \\
&\quad \texttt{USDC\_transferFrom(user, curveDeposit, uamounts,...); } &\\
&\quad \texttt{USDC\_approve(compCUSDC, uamounts, curveDeposit);} &\\
&\quad \texttt{cUSDC\_mint(uamounts,curveDeposit);} &\\
&\quad \texttt{tau\{cAmounts[USDC] = cUSDC\_accountTokens[curveDeposit];\}} &\\
&\quad \texttt{cUSDC\_approve(curveSwap, cAmounts[USDC], curveDeposit);} &\\
&\quad \texttt{Curve\_swap\_addLiquidity(cAmounts, min\_mint\_amount, sender);} &\\
&\quad \texttt{cCrv\_transfer(user, cCrv\_mintAmounts, curveDeposit)\};} &
\end{flalign*}
\caption{CSP\# definition of \texttt{add\_liquidity()} function accepting USDC.}
\label{fig:add-csp}
\end{subfigure}
\caption{Implementations of \texttt{add\_liquidity()} function in Curve Compound pool.}
\label{fig:curve-func}
\end{figure}

\begin{figure}[h]
\centering
\begin{subfigure}[b]{\textwidth}
\centering
\includegraphics[width=\textwidth]{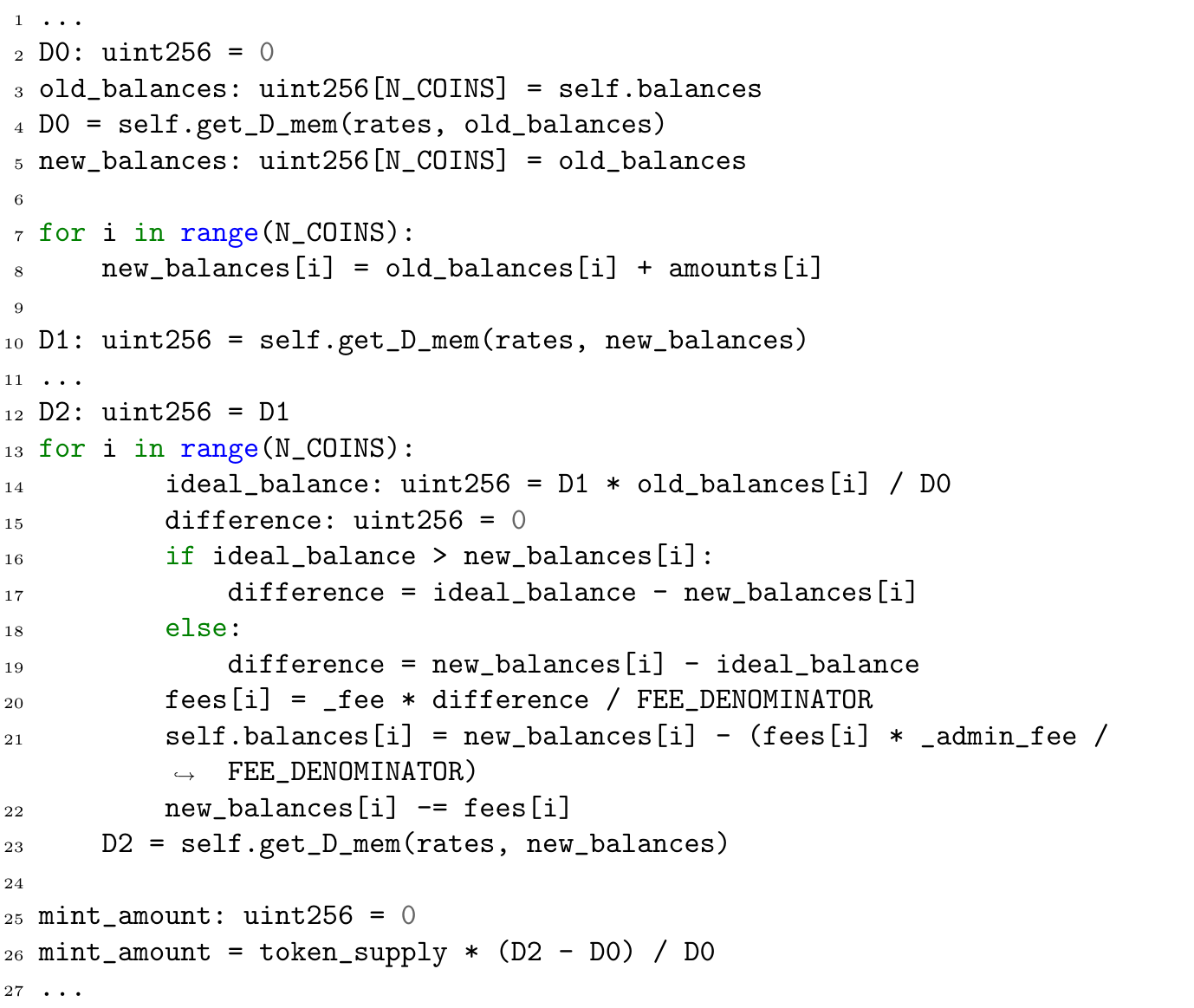}
\caption{Vyper implementation of calculations in Curve pool.}
\label{fig:vyper-add-swap}
\end{subfigure}
\begin{subfigure}[b]{\textwidth}
\centering
\includegraphics[width=\textwidth]{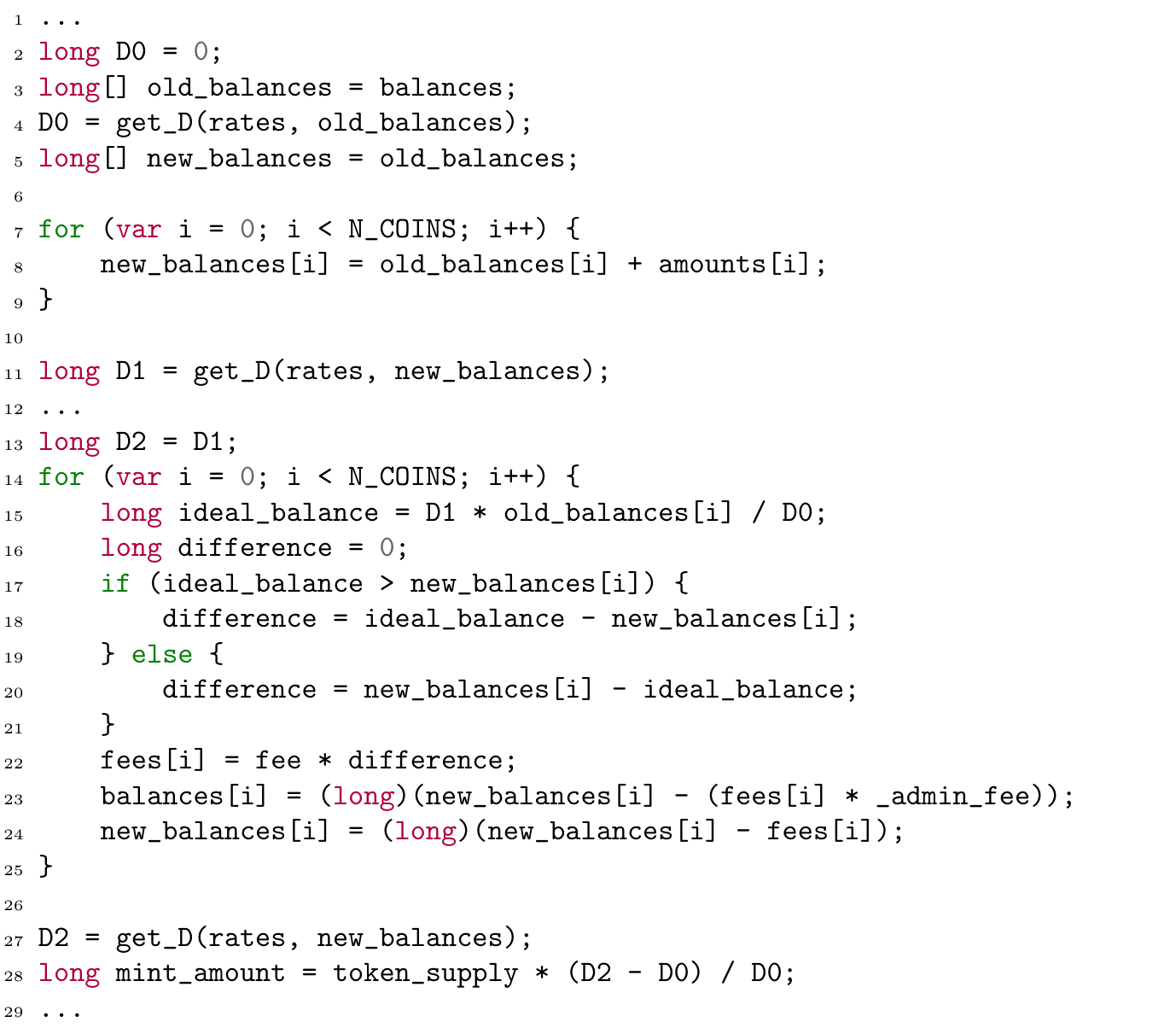}
\caption{C\# implementation of calculations in Curve pool.}
\label{fig:csharp-add-swap}
\end{subfigure}
\caption{Implementations of mathematical computations in Curve.}
\label{fig:curve-swap-func}
\end{figure}

\Cref{fig:curve-func} demonstrates the simplified Vyper code of the function that implements adding
liquidity (\texttt{add\_liquidity()}) to the Curve Compound
pool\footnote{https://etherscan.io/address/0xeb21209ae4c2c9ff2a86aca31e123764a3b6bc06}
and its definition in CSP\#.
The function~(\cref{fig:add-vyper}) and the corresponding process~(\cref{fig:add-csp}) perform transfers
of liquidity (USDC) and pool (cCrv)\footnote{https://etherscan.io/address/0x845838DF265Dcd2c412A1Dc9e959c7d08537f8a2}
tokens as described in~\cref{fig:flow}. The mathematical computation of the number of
pool tokens to mint is partially implemented in C\#. Vyper and C\# code that implement the calculation
are shown in~\cref{fig:curve-swap-func}.

\end{document}